\documentstyle[prl,aps,floats]{revtex}
\def\be{\begin{eqnarray}}
\def\ee{\end{eqnarray}}
\def\ben{\begin{equation}}
\def\een{\end{equation}}

\def\calM{{\cal M}}

\newcommand{\e}{{\mbox{e}}}
\def\mN{m_N}
\def\del{\partial}

\def\vr{{\vec r}}
\def\vq{{\vec q}}

\def\vp{{\vec p}}

\def\vbp{{\vec {\bar p}}}

\def\vs{{\vec \sigma}}

\def\roughly#1{\mathrel{\raise.3ex\hbox{$#1$\kern-.75em%
\lower1ex\hbox{$\sim$}}}}

\def\fm{{\mbox{fm}}}
\def\MeV{{\mbox{MeV}}}
\def\GeV{{\mbox{GeV}}}

\def\ve{\varepsilon}

\def\nlo#1{\mbox{N$^{#1}$LO}}

\def\dR{{\hat d}^R}

\begin{document}

\twocolumn[%
\hsize\textwidth\columnwidth\hsize\csname@twocolumnfalse\endcsname

\renewcommand{\thefootnote}{\fnsymbol{footnote}}
\setcounter{footnote}{0}

\title{%
Parameter-Free Calculation of the Solar Proton Fusion Rate
in Effective Field Theory}

\author{%
{\bf T.-S. Park}$^{(a)}$, {\bf  L.E. Marcucci}$^{(b,c)}$, {\bf R.
Schiavilla}$^{(d,e)}$, {\bf M. Viviani}$^{(c,b)}$, {\bf A.
Kievsky}$^{(c,b)}$, {\bf S. Rosati}$^{(b,c)}$
\\
{\bf K. Kubodera}$^{(a)}$, {\bf D.-P. Min}$^{(f)}$, and {\bf M.
Rho}$^{(f,g,h)}$ }
\address{%
(a) Department of Physics and Astronomy,
University of South Carolina, Columbia, SC 29208, USA\\
(b) Department of Physics, University of Pisa, I-56100 Pisa, Italy\\
(c) INFN, Sezione di Pisa, I-56100 Pisa, Italy\\
(d) Department of Physics, Old Dominion University,
         Norfolk, Virginia 23529, USA \\
(e) Jefferson Lab, Newport News, Virginia 23606, USA \\
(f) School of Physics and Center for Theoretical Physics,,
Seoul National University, Seoul 151-742, Korea\\
(g) Service de Physique Th\'{e}orique, CEA  Saclay, \it
91191 Gif-sur-Yvette Cedex, France\\
(h) Institute of Physics and Applied Physics, Yonsei University,
Seoul 120-749, Korea
}

\maketitle

\centerline{(\today)}

\begin{abstract}
Spurred by the recent complete determination of
the weak currents in two-nucleon systems up to ${\cal O}(Q^3)$
in heavy-baryon chiral perturbation theory,
we carry out a parameter-free calculation
of the solar proton fusion rate
in an effective field theory that {\it combines}
the merits of the standard nuclear physics method and
systematic chiral expansion.
Using the tritium $\beta$-decay rate as an input
to fix the only unknown parameter in the effective Lagrangian,
we can evaluate with drastically improved precision
the ratio of the two-body contribution
to the well established one-body contribution;
the ratio is determined to be $(0.86\pm 0.05)\ \%$.
This result is essentially independent of the cutoff
parameter for a wide range of its variation
($500\ \MeV \le \Lambda \le 800\ \MeV$),
a feature that substantiates the consistency
of the calculation.
\end{abstract}

 \vskip1pc]

\renewcommand{\thefootnote}{\arabic{footnote}}
\setcounter{footnote}{0}

The principal role of effective field theories (EFT in short) in
nuclear physics is two-fold. One is to describe nuclear dynamics
starting from a ``first-principle" theory anchored on QCD. As is
well known, the standard nuclear physics approach (SNPA) based on
phenomenological potentials has been enormously successful
\cite{cs98}; meanwhile, there is growing interest in establishing
the foundation of SNPA, identifying nuclear physics as a {\it bona fide}
element of the fundamental Standard Model.  The second role of EFT
in nuclear physics is to make precise model-independent
predictions for nuclear observables with quantitative estimates of
uncertainties attached to them.  This second goal is particularly
important in providing nuclear physics input needed in
astrophysics. In focusing on the second objective, it has been
emphasized in a series of recent articles~\cite{PKMR} that a very
promising approach is to combine the highly developed SNPA
with an EFT based on chiral dynamics of QCD. This approach is
intended to take full advantage of the extremely high accuracy of
the wave functions achieved in SNPA while securing a good control
of the transition operators 
via systematic chiral expansion.
(See Ref.~\cite{bc01} for an another EFT calculation on
$pp$ fusion.)
Such an approach--which is close in spirit to Weinberg's original
scheme~\cite{weinberg} based on the chiral expansion of
``irreducible terms"--has been found to have an amazing
predictive power for the $n + p\rightarrow d+\gamma$
process~\cite{np,npp}.

In this paper we apply the same strategy
to the solar proton fusion process
\be
 p+p\rightarrow d + e^+ +\nu_e\>\>.\label{pp}
\ee
The process (\ref{pp}) was previously analyzed
in Ref. \cite{pp}
(hereafter referred to as PKMR98)
by four of the authors
in heavy-baryon chiral perturbation theory (HB$\chi$PT).
The calculation was carried out
up to ${\cal O}(Q^3)$ in chiral order,
viz., next-to-next-to-next-to-leading order (\nlo3).
At \nlo3, two-body meson-exchange currents (MEC)
begin to contribute,
and there appears one unknown parameter
in the chiral Lagrangian contributing to the MEC.
This unknown constant, called $\dR$ in \cite{pp},
represents the strength of a contact interaction.
One intuitively expects that zero-ranged terms of this sort
are suppressed by hard-core correlations in the wave functions,
and hence it is a common practice
to drop their contributions altogether.
This approximation--referred to as the hard-core
cutoff scheme (HCCS)--can indeed be justified in cases
where the ``chiral filter mechanism"
associated with pion dominance holds~\cite{np}
(see \cite{PKMR} for details).
It turns out, however,
that the ``chiral filter mechanism"
does not apply to the Gamow-Teller (GT) matrix element
that gives the dominant contribution to the process (\ref{pp}).
Thus there is no good reason to argue away
the contact term contribution in this case.
For practical reasons, however,
the $\dR$-term contribution
was simply ignored in \cite{pp}
by invoking the HCCS and a ``naturalness" argument.
Let $\delta_{\rm 2B}$ stand for the ratio
of the contribution of the two-body MEC
to that of the one-body current.
In \cite{pp}, $\delta_{\rm 2B}$ for the case
without the contact term is found to be
$\delta_{\rm 2B}= (4.0 \pm 0.5)\ \%$,
where the ``errors" reflect changes in $\delta_{\rm 2B}$
as the hard core radius varies within the range,
$0.55\ \fm \le r_C \le 0.80\ \fm$.
To accommodate uncertainty associated with the contact term
which was dropped by {\it fiat},
the total MEC contribution was assigned
$\sim$ 100 \% error.  With this large uncertainty attached,
an EFT prediction on the $pp$ fusion rate reported in \cite{pp}
was unable to corroborate or exclude the recent results
of SNPA calculations \cite{schiavilla},
according to which $\delta_{\rm 2B}=0.5 \sim 0.8$ \%.

We argue here that the situation can be improved dramatically.
The crucial point is that exactly the same combination
of counter terms that defines the constant $\dR$
enters the GT matrix elements
that feature in $pp$ fusion, tritium $\beta$-decay,
the $hep$ process, $\mu$-capture on a deuteron,
and $\nu$--$d$ scattering.
This means that, if the value of $\dR$ can be fixed
using one of these processes, then
it is possible to make a totally parameter-free prediction
for the GT matrix elements
of the other processes.
The existence of accurate experimental data
for the tritium $\beta$-decay rate, $\Gamma_\beta$,
indeed allows us to carry out this program;
here we are specifically interested
in the model-independent determination
of the $pp$ fusion rate.
The availability of extremely well tested,
realistic wave functions for the $A$=3 nuclear systems
enables us to eliminate ambiguities
related to nuclear many-body problems.
A particularly favorable aspect pertaining to
tritium $\beta$-decay as well as $pp$ fusion
is that both of them are dominated by the GT operator
to a very high degree.
In order for our result to be physically acceptable,
however, its cutoff dependence must be under control.
In our scheme, for each value of $\Lambda \sim 1/r_C$
that defines the energy/momentum cutoff scale of EFT,
$\dR$ is determined to reproduce $\Gamma_\beta$;
thus $\dR$ is a function of $\Lambda$.
The premise of EFT is that,
even if $\dR$ itself is $\Lambda$-dependent,
physical observables (in our case the $pp$-fusion rate)
should be independent of $\Lambda$
as required by renormalization-group invariance.
We shall show below that our method
indeed gives an essentially
$\Lambda$-independent result,
$\delta_{\rm 2B} \simeq (0.86\pm 0.05)$ \%.
With this refined estimate of the two-body correction
to the well established one-body contribution,
we are in a position to make a parameter-free prediction
for the astrophysical $S$ factor for $pp$ fusion
with drastically improved precision.

It is worth emphasizing that the above EFT prediction
for $\delta_{\rm 2B}$ is in line with the latest SNPA
results obtained in Ref.~\cite{schiavilla} (and mentioned earlier).  
There too, the short range behavior of the axial MEC was
constrained by reproducing the GT matrix element of tritium
$\beta$-decay.  The inherent model dependence of such a procedure
within the SNPA context was shown to be very weak simply because
at small inter-particle separations, where
MEC contributions are largest, the pair wave functions in
different nuclei are similar in shape and differ only by a scale
factor~\cite{forest96}.  As a consequence, the ratios of GT and $p$$p$-capture
matrix elements of different two-body current terms are nearly the same,
and therefore knowledge of their sum in the GT matrix element
is sufficient to predict their sum in the
$p$$p$-capture matrix element~\cite{schiavilla}.

For the solar $pp$-fusion process and tritium $\beta$-decay,
it is sufficient to consider the limit
in which the nuclear system receives no momentum transfer.
In this limit, 
the one- and two-body axial-vector currents 
of the $a$-th isospin component read
(in the conventions used in PKMR98)
\be
{\vec A}^a_{\rm 1B} &=& g_A \sum_{l=1,2}
\frac{\tau^a_l}{2} \left[
 \vs_l + \frac{\vbp_l\, \vs_l
 \cdot \vbp_l - \vs_l \, {\bar p}_l^2}{2 m_N^2}
 \right],
 \label{1-body}
\\
{\vec A}_{{\rm 2B}}^a &=&
- \frac{g_A}{2 \mN f_\pi^2}\, \frac{1}{m_\pi^2 + q^2}
 \Bigg[ - \frac{i}{2} \tau^a_\times\,\vp\,\,
     (\vs_1-\vs_2)\cdot\vq
 \nonumber \\
 &&+ 2 \,\hat c_3\, \vq \,\, \vq\cdot
  (\tau_1^a \vs_1 +\tau_2^a \vs_2)
\nonumber \\
 &&+ \left(\hat c_4
  + \frac14\right) \tau^a_\times\,
 \vq \times \left[ \vs_\times \times \vq\, \right]
 \frac{}{}\Bigg]
\nonumber \\
 &&- \frac{g_A}{m_N f_\pi^2} \left[
  \hat d_1 (\tau_1^a \vs_1 + \tau_2^a \vs_2)
 + \hat d_2 \tau^a_\times \vs_\times
 \right] \>\>,\label{2-body}
\ee 
with $\vp \equiv (\vbp_1 - \vbp_2)/2$, $\vbp_l\equiv
(\vp_l+ \vp_l^{\,\prime})/2$, 
$\tau^a_\times\equiv (\tau_1\times\tau_2)^a$
and similarly for $\vs_\times$.
Eq.(\ref{1-body}) includes a $1/m_N^2$
correction term, which was ignored in PKMR98,
and which is found to decrease the one-body matrix element by
$0.13$ \%.
In terms of $\Lambda_{pp}$
defined in Ref.~\cite{KB94}~\footnote{The subscript $pp$ has
been added here to avoid confusion with the cutoff parameter
$\Lambda$. 
} we have
\be
\Lambda_{pp}^2&\equiv& 
\frac{|a^C|^2 \gamma^3}{2}\, A_S^2\, \calM_{\rm 1B}^2 
= 6.91 \ee 
for the
central value. This should be compared with $6.93$ obtained in
\cite{pp}.

The values of $\hat c$'s in eq.(\ref{2-body}) 
have been determined by $\pi$-$N$ data \cite{csTREE}:
${\hat c}_3= -3.66 \pm 0.08$
and ${\hat c}_4= 2.11 \pm 0.08$.
The two constants,
$\hat d_{1}$ and $\hat d_2$, remain to be fixed
but,
thanks to Fermi-Dirac statistics,
only one combination turns out to be relevant
to the $pp$ fusion process\footnote{
There was a sign error
in the corresponding expression in PKMR98;
as corrected here, the $\hat d_2$ term
should have the positive sign.
This error, however, does not affect the numerical results of PKMR98,
since the resulting error gets absorbed into
the definition of $\dR$.}:
\be
 \dR\equiv \hat d_1 +2 \hat d_2 +
\frac13 \hat c_3
 + \frac23 \hat c_4 + \frac16 \>\>.
\label{dr}
\ee
Most fortunately, as mentioned, the same combination enters
into tritium $\beta$-decay, the $hep$ process,
$\mu$-capture on deuteron, and $\nu$--$d$ scattering.
We choose here to determine $\dR$ using 
the tritium $\beta$-decay rate, $\Gamma_\beta$, for which there 
are accurate experimental results.
To this end, we calculate $\Gamma_\beta$
from the matrix elements of the current operators
in Eqs.(\ref{1-body}) and (\ref{2-body})
evaluated for appropriate $A$=3 nuclear wave functions. 
We employ here the wave functions
obtained in Ref.~\cite{roccoetal}
using the correlated-hyperspherical-harmonics (CHH) method.
It is obviously important
to maintain consistency between
the treatments of the $A$=3 and $A$=2 systems.
In our case the same Argonne $v_{18}$ (AV18) potential~\cite{av18}  
is used (with the addition of the Urbana-IX three-nucleon 
potential~\cite{uix} for the $A$=3 nuclei).
Furthermore, we apply the same regularization method
to both $A$=2 and $A$=3 systems
to control short-range physics in a consistent manner.
Specifically, in performing Fourier transformation
to derive the $r$-space representation of
a transition operator,
we use the Gaussian regularization.
This is equivalent
to replacing the delta and Yukawa functions
with the regulated ones,
\be
\left( \delta_\Lambda^{(3)}(r),\,
y_{0\Lambda}^\pi(r) \right) 
 \equiv
 \int \!\!\frac{d^3 q}{(2\pi)^3}\,
  S_\Lambda^2(q^2)\, \e^{ i \vq\cdot \vr} 
  \left(1,\,\frac{1}{q^2 + m_\pi^2}\right) ,
\nonumber\ee
$y_{0\Lambda}^\pi(r) \equiv
- r \frac{\del}{\del r} y_{0\Lambda}^\pi(r)$
and
$y_{2\Lambda}^\pi(r) \equiv \frac{1}{m_\pi^2}
  r \frac{\del}{\del r} \frac{1}{r}
  \frac{\del}{\del r} y_{0\Lambda}^\pi(r)$,
where $S_\Lambda(q^2)$ is defined as 
\be S_\Lambda(q^2) =
\exp\left(-\frac{q^2}{2\Lambda^2}\right). \label{regulator} \ee
The cutoff parameter $\Lambda$ characterizes the energy-momentum
scale of our EFT.  The properly regularized two-body matrix
elements read 
\be \calM_{\rm 2B} &=&
 \frac{2}{\mN f_\pi^2}
 \int_0^\infty\! dr\,  \left\{ \frac{}{} \right.
\nonumber \\
&&
 \frac{m_\pi^2}{3} 
\left(\hat c_3 + 2 \hat c_4 + \frac12\right)
  y_{0\Lambda}^\pi(r)\, u_d(r)\,  u_{pp}(r) 
\nonumber\\
&-& \sqrt{2}
\frac{m_\pi^2}{3}
 \left(\hat c_3 - \hat c_4 - \frac14\right)
 y_{2\Lambda}^\pi(r)\, w_d(r) \, u_{pp}(r)
\nonumber \\
&+& %\frac{m_\pi}{r}
 \frac{y_{1\Lambda}^\pi(r)}{12 r} \Bigg[
 \left[ u_d(r)-\sqrt2 w_d(r)\right]  u_{pp}'(r) \nonumber \\
&-& \left[ u_d'(r)-\sqrt2 w_d'(r)\right] u_{pp}(r)
+ \frac{3\sqrt2}{r} w_d(r) u_{pp}(r)
\Bigg] \nonumber \\
&-&\left. \dR \delta_\Lambda^{(3)}(r)\, u_d(r) u_{pp}(r) 
\frac{}{}\right\},
\label{calM2Bdelta}
\ee
where $u_d(r)$ and $w_d(r)$ are the S- and D-wave components
of the deuteron wave function, and $u_{pp}(r)$ is the $^1$S$_0$ $pp$ scattering
wave (at zero relative energy).

As mentioned, we can associate the momentum cutoff parameter
$\Lambda$ in Eq.(\ref{regulator}) to the cutoff scale of our EFT.
We let $\Lambda$ vary within a certain reasonable range and, for a
given value of $\Lambda$, we deduce the value of $\hat{d}^R$ that
reproduces $\Gamma_\beta$. With $\hat{d}_R$ so determined, we
calculate the two-body matrix element $\calM_{\rm 2B}$. The
results are given for three representative values of $\Lambda$ in
Table 1. 
The table indicates
that, although the value of $\dR$ is sensitive to $\Lambda$,
$\calM_{\rm 2B}$ is amazingly stable against the variation of
$\Lambda$ over a wide range. In view of this high stability, we
believe we are on the conservative side in adopting the estimate
$\calM_{\rm 2B}= (0.039 \sim 0.044)\ \fm$. Since the leading
single-particle term is independent of $\Lambda$, the total
amplitude $\calM = \calM_{\rm 1B}+\calM_{\rm 2B}$ is
$\Lambda$-independent to the same degree as $\calM_{\rm 2B}$. The
$\Lambda$-independence of the physical quantity $\calM$, which is
in conformity with the general {\it tenet} of EFT, is a crucial feature
of the result in our present study. The relative strength of the
two-body contribution as compared with the one-body contribution is 
\be 
\delta_{\rm 2B} \equiv \frac{\calM_{\rm 2B}}{\calM_{\rm 1B}} 
= (0.86 \pm 0.05)\ \%. 
\label{delta2B-new} 
\ee 
We remark that
the central value of $\delta_{\rm 2B}$ here is considerably
smaller than the corresponding value, $\delta_{\rm 2B}= 4$ \%, in
PKMR98. Furthermore, the uncertainty of $\pm$0.05 \% in
Eq.(\ref{delta2B-new}) is drastically smaller than the
corresponding figure, $\pm$4 \%, in PKMR98.

\begin{table}[b]
\begin{center}
\begin{tabular}{|c|c|l|}
$\Lambda$ (MeV) & $\dR$ & $\calM_{\rm 2B}$ (fm) \\
\hline
500 & $1.00 \pm 0.07$ &
$0.076 - 0.035\ \dR \simeq 0.041 \pm 0.002 $\\
\hline
600 & $1.78 \pm 0.08$ &
$0.097 - 0.031\ \dR \simeq 0.042 \pm 0.002$
\\ \hline
800 & $3.90 \pm 0.10$ &
$0.129 - 0.022\ \dR \simeq 0.042 \pm 0.002$
\\ 
\end{tabular}
\caption{\protect The strength $\dR$
of the contact term and the two-body
GT matrix element, $\calM_{\rm 2B}$,
calculated for representative values of $\Lambda$.}
\end{center}
\end{table}

We now turn to the threshold $S$ factor, which is a
key input for the solar model. 
Adopting the most recent values
$g_A= 1.2670 \pm 0.0035$
\cite{PDG} and $G_V=(1.14939 \pm 0.00065)\times 10^{-5}\
\GeV^{-2}$ \cite{hardy},
we have
\be S_{pp}(0) &=& 
3.94\ \left(\frac{1+\delta_{2B}}{1.01}\right)^2
\left(\frac{g_A}{1.2670}\right)^2
\left(\frac{\Lambda_{pp}^2}{6.91}\right)^2
\nonumber \\
&=& 
 3.94\ (1 \pm 0.15\ \% \pm 0.10\ \% \pm \ve) 
    \label{S-factor}
\ee
in units of $10^{-25}\ \mbox{MeV-barn}$.
Here the first error is due to uncertainties
in the input parameters in the one-body part,
while the second error represents the uncertainties
in the two-body part;
$\ve(\approx 0.1\ \%)$ denotes possible uncertainties
due to higher chiral order contributions (see below).
To make a formally rigorous assessment of $\ve$, we must evaluate
loop corrections and higher-order counter terms. 
Although an N$^4$LO calculation would not involve any new 
unknown parameters, it is a non-trivial task.  
Furthermore, loop corrections necessitate a more
elaborate regularization scheme since the naive cutoff
regularization used here violates chiral symmetry at loop orders.
(This difficulty, however, is not insurmountable.) These formal
problems set aside, it seems reasonable to assess $\ve$ as
follows. We first recall that both tritium $\beta$-decay and solar
$pp$ fusion are dominated by the one-body GT matrix elements, the
evaluation of which is extremely well controlled from the SNPA as
well as EFT points of view. Therefore,
the precision of our calculation is
governed by the reliability of estimation of small corrections to
the dominant one-body GT contribution. Now, we have seen that the
results of the present \nlo3 calculation nicely fit into the
picture expected from the general {\it tenet} of EFT: (i) the \nlo3
contributions are indeed much smaller than the leading order term;
(ii) the physical transition amplitude $\calM$ does not depend on
the cutoff parameter.  Although these features do not constitute a
formal proof of the convergence of the chiral expansion used here, it
is {\it extremely unlikely} that higher order contributions be so
large as to completely upset the physically reasonable behavior
observed in the \nlo3 calculation. It should therefore be safe to
assign to $\ve$ uncertainty comparable to the error estimate for
the two-body part in Eq.(\ref{S-factor}); viz., $\ve \approx
0.1$ \%. 
In this connection we remark that
an axial three-body MEC contribution
to the $^3$H GT matrix element
was calculated explicitly in SNPA~\cite{roccoetal}
and found to be negligible relative 
to the leading two-body mechanisms.
This feature is consistent with the above argument
since, in the context of EFT, 
the three-body MEC represents a higher-order effect
subsumed in $``\ve"$ in Eq.(\ref{S-factor}).

Apart from the  notable numerical differences between the present
work and PKMR98, it is important to point out that short-range
physics is much better controlled here. In the conventional
treatment of MEC, one derives the coordinate space representation
of a MEC operator by applying ordinary Fourier transformation
(with no restriction on the range of the momentum variable) to the
amplitude obtained in momentum space; this corresponds to setting
$\Lambda=\infty$ in Eq.(\ref{regulator}). In PKMR98, where this
familiar method is adopted, the $\dR$ term appears in the
zero-range form, $\dR\delta(r)$. Since there were no known
methods to fix the parameter $\dR$ (except for rough estimates
based on the naturalness argument), PKMR98 chose to introduce
short-range repulsive correlation with hard-core radius $r_C$ and
eliminate the $\dR\delta(r)$ term {\it by hand}. The
remaining finite-range terms were evaluated as functions of $r_C$.
$\calM_{\rm 2B}$ calculated this way exhibited substantial
$r_C$-dependence, indicating that short-range physics was not well
controlled; if $1/r_C$ is interpreted as a typical value of the
momentum cutoff $\Lambda$ in EFT, the results in PKMR98 imply that the
theory has significant dependence on $\Lambda$.  Inclusion of the
$\dR$ term, with its strength renormalized as described here,
eliminates this undesirable $\Lambda$-dependence to a satisfactory
degree.

As mentioned, the methodology employed in this work can be
profitably used for the other related processes. An application to
the $hep$ process will be reported in a separate article
\cite{coll}.

TSP and KK thank S. Ando and F. Myhrer for discussions.
The work of TSP and KK is supported in part
by the U.S. National Science Foundation,
Grant No.~PHY-9900756 and No.~INT-9730847.
The work of RS is supported by DOE contract No.~DE-AC05-84ER40150
under which the Southeastern Universities Research Association (SURA)
operates the Thomas Jefferson National Accelerator Facility.
The work of DPM is supported in part by KOSEF Grant
1999-2-111-005-5 and KSF Grant 2000-015-DP0072.
MR acknowledges the hospitality
of the Physics Departments of SNU and Yonsei University,
where his work was partially supported by Brain Korea21 in 2001.
Some of the calculations were made possible by grants
of computing time from the National Energy Research Supercomputer
Center in Livermore.

\end{document}